\documentstyle [12pt] {article}
\title{Analytical theory of multipass crystal extraction}

\author{Valery Biryukov\thanks{E-mail:  biryukov@mx.ihep.su}
\\ {\small Institute for High Energy Physics}   \\
{\small Protvino, 142284 Moscow Region, Russia}
\\ \vspace{10mm} Published in EPAC 1998 Proceedings, pp.2091-2093}

\date{June 1998}
\begin{document}
\maketitle

\begin{abstract}
An analytical theory for the efficiency of particle
extraction from an accelerator by means of a bent crystal
is proposed. The theory agrees with all the measurements
performed in the broad energy range of 14 to 900 GeV,
where the efficiency range also spans over two decades,
from $\sim$0.3\% to $\sim$30\%.
Possibilities for crystal extraction from sub-GeV accelerators
and from muon colliders are discussed.
\end{abstract}

\section{Introduction}
Crystal extraction experiments have greatly progressed
at high energy accelerators in recent years [1,2].
The experimental data are in good agreement with predictions
from the detailed
Monte-Carlo simulations [3,4].
Although the transmission of particles by a bent crystal
can be described analytically with good accuracy,
the process of extraction involves essentially multiple encounters
of circulating particles with the crystal and many turns in the
accelerator, and therefore its efficiency cannot be scaled easily,
for instance with energy.
An analytical theory of multipass crystal extraction
would be highly helpful in understanding the existing
experimental results, in extrapolation to future applications,
and in optimization.
Below we derive an analytical formula for the crystal
extraction efficiency.

\section{Theory}
Suppose that a beam with divergence $\sigma$, Gaussian distribution,
is aligned to the crystal planes. Then as many as
$(2\theta_c/\sqrt{2\pi}\sigma ) (\pi x_c/2d_p)$
particles get channeled in the initial straight part of the crystal.
Here $\theta_c$ stands for the critical angle of channeling,
$d_p$ the interplanar spacing, $x_c\approx d_p/2-a_{TF}$
the critical distance, $a_{TF}$ being the Thomas-Fermi screening
distance.

We shall first consider the more typical case,
where particles first come to
the crystal with nearly zero divergence, due to very small impact
parameters.
As experiments indicate [1,2],
in this first passage the channeling
is suppressed, apparently due to the poor quality of
the crystal structure near surface.
In our model we assume that the first passage of particle through the
crystal is always "inefficient",
i.e. there is no channeling, but there is
scattering and possibly nuclear interactions.

After some turns in the accelerator ring, the scattered particles come
to the crystal with rms divergence as defined by scattering in the
first pass:
$\sigma_1 = (E_s/pv)(L/L_R)^{1/2}$,
where $E_s$=13.6 MeV, $L$ is the crystal length, $L_R$ the radiation length,
$pv$ the particle momentum times velocity.
In a real experiment, $\sigma_1$ may be affected by betatron oscillations,
and also by the fact that in the first passage the particle may enter
the bent crystal quite near its surface and hence leave the crystal
before crossing its full length $L$.
These complications are to be taken into account in
the detailed Monte Carlo simulations [3,4], as well as the
apertures, etc.
However, our objective is to derive a simple analytical theory
which includes only the basic physical parameters of crystal extraction
process, and to see how far it goes.
We assume then that any particle always crosses the full crystal
length; that pass 1 is like through an amorphous matter but any
further pass is like through a crystalline matter;
that there are no aperture restrictions; and that the particles
interact only with the crystal not a holder.

After $k$ passes the divergence is $\sigma_k=k^{1/2}\sigma_1$.
The number of particles lost in nuclear interactions is
1$-\exp(-kL/L_N)$ after $k$ passes; $L_N$ is the interaction length.
In what follows we shall first assume that
the crystal extraction efficiency is substantially smaller than
100\% (which has actually been the case so far), i.e. the circulating
particles are removed from the ring predominantly through the
nuclear interactions, not through channeling.

That pulled together, we obtain the multipass channeling
efficiency by summation over $k$ passes:
\begin{equation}
F_C=\left(\frac{\pi}{2}\right)^{1/2}\frac{\theta_cx_c}{\sigma_1d_p}
   \times \Sigma(L/L_N)
\end{equation}
where
\begin{equation}
\Sigma(L/L_N)= \Sigma_{k=1}^{\infty} k^{-1/2}\exp(-kL/L_N)
\end{equation}
may be called a "multiplicity factor" as it just tells how much
the single-pass efficiency is amplified in multipasses.

A fraction of channeled particles is to be lost along the bent
crystal due to scattering processes and centripetal effects.
The transmission factor for the channeled particles in a bent crystal
we denote as $T$. Then the multipass extraction efficiency is
\begin{equation}
F_E=F_C\times T=
  \left(\frac{\pi}{2}\right)^{1/2}\frac{\theta_cx_c}{\sigma_1d_p}
    \times \Sigma(L/L_N) \times T
\end{equation}
We shall use an analytical approximation (as used also in [6])
for silicon
\begin{equation}
  T = (1-p/3R)^2 \exp\left(-\frac{L}{L_d(1-p/3R)^2}\right),
\end{equation}
where $p$ is in GeV/c, and $R$ is in cm;
$L_d$ is dechanneling length for a straight crystal.
The first factor in $T$ describes a centripetal dechanneling.
E.g., at $pv/R$=0.75 GeV/cm (which is close to the highest
values used in extraction)
our approximation gives $(1-p/3R)^2$=0.563
whereas Forster et al. [7] measured 0.568$\pm$0.027.
The dechanneling length $L_d$ we describe by the theoretical formula[9].

To have all formulas explicit,
we give an analytical expression for the sum (2),
for $L\ll L_N$:
\begin{equation}
\Sigma(L/L_N) \simeq (\pi L_N/L)^{1/2}-1.5
\end{equation}

\section{Experimental check}
Let us check the theory, first against the CERN SPS data [10]
where the crystal
extraction efficiency was measured at 14, 120, and 270 GeV,
making use of the same 4-cm long Si(110) crystal, deflecting
at 8.5 mrad.
The crystal had 3-cm long bent part with two 5-mm straight ends,
having in the center $pv/R$=(0.34 GeV/cm)$\times$($pv$/120 GeV).
We take $\theta_c$=13.8 $\mu$rad$\times$(120 GeV$/pv)^{1/2}$
(as used by the authors of Ref.[10]).
The dechanneling length for a straight crystal is taken as
0.569$\times$270=154 mm at 270 GeV,
0.603$\times$120=72.4 mm at 120 GeV, and
0.718$\times$14=10.1 mm at 14 GeV.
This length is reduced in a bent crystal by a factor
of (1$-p/3R)^2$, Eq.(4).
Only the dechanneling over 35 mm is taken into account, as the last
5-mm end is unbent.
Table 1 shows good agreement of theory with measurements.
\begin{table}[htb]
\begin{center}
\caption{Extraction efficiencies (\%) from the SPS experiment,
 Eq.(3), and detailed simulations [11].}
\begin{tabular}{cccc}
 & & & \\
                $pv$(GeV) &    SPS     &  Eq.(3) & Monte Carlo \\
\hline
 & & & \\
                14 &   0.55$\pm$0.30&  0.30 & 0.35$\pm$0.07 \\
               120 &   15.1$\pm$1.2 & 13.5 & 13.9$\pm$0.6  \\
               270 &   18.6$\pm$2.7 & 17.6 & 17.8$\pm$0.6
\end{tabular}
\end{center}
\end{table}
\begin{figure}[htb]
\begin{center}
\setlength{\unitlength}{.5mm}
\begin{picture}(110,80)(5,-10)
\thicklines
\linethickness{.25mm}
\put(    40.,27.)  {\circle {3}}
\put(    90.,35.20)  {\circle {3}}
\put(    4.7,.6)  {\circle {3}}
\put(    0.,0.)  {\circle {3}}
\put(    8.3,3.2)  {\circle {3}}
\put(    13.3,8.)  {\circle {3}}
\put(    23.3,17.)  {\circle {3}}
\put(    33.3,23.8)  {\circle {3}}
\put(    50.,30.8)  {\circle {3}}
\put(    80.,34.6)  {\circle {3}}
\put(    66.7,34.1)  {\circle {3}}
\put(    117.,32.8)  {\circle {3}}
\put(    133.,30.)  {\circle {3}}
\put(    100.,34.6)  {\circle {3}}
\put(    140.,26.8)  {\circle {3}}

\put(40,27.8) {\line(0,1){4.8}}
\put(39,30.2) {\line(1,0){2}}
\put(90,31.8) {\line(0,1){10.8}}
\put(89,37.2) {\line(1,0){2}}
\put(4.7,.5) {\line(0,1){1.2}}
\put(3.7,1.1) {\line(1,0){2}}

\put(0,0) {\line(1,0){140}}
\put(0,0) {\line(0,1){60}}
\put(0,60) {\line(1,0){140}}
\put(140,0){\line(0,1){60}}
\multiput(16.667,0)(16.667,0){8}{\line(0,-1){2}}
\multiput(0,20)(0,20){3}{\line(1,0){2}}
\multiput(0,4)(0,4){15}{\line(1,0){1.4}}
\put(33,-7){\makebox(1,1)[b]{100}}
\put(66,-7){\makebox(1,1)[b]{200}}
\put(100,-7){\makebox(1,1)[b]{300}}
\put(133,-7){\makebox(1,1)[b]{400}}
\put(-11,20){\makebox(1,.5)[l]{10}}
\put(-11,40){\makebox(1,.5)[l]{20}}

\put(-17,50){\large F(\%)}
\put(73,-14){\large $p$ (GeV/c)}

\end{picture}
\end{center}
\caption{
The SPS extraction efficiency as a function of
momentum $p$. The curve (o) is for Eq.(3),
the crosses at 14, 120 and 270 GeV/c are for the SPS experiment.
}
\end{figure}
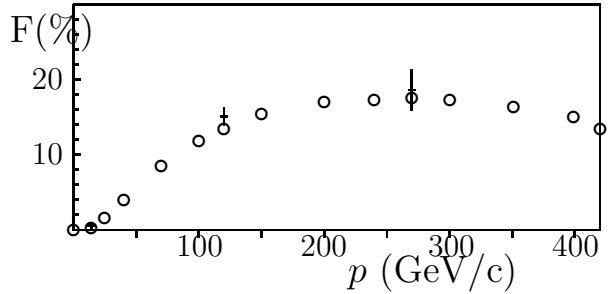

The Tevatron extraction experiment at 900 GeV provides
another check at a substantially higher value of efficiency.
Here a slight modification of the formulas is needed to
account for the non-zero starting divergence, namely
$\sigma_0$=11.5 $\mu$rad (rms).
This results in the change in Eq.(2):
\begin{equation}
\Sigma(L/L_N)= \Sigma_{k=1}^{\infty}
  (k+\sigma_0^2/\sigma_1^2)^{-1/2}\exp(-kL/L_N)
\end{equation}
Since in this experiment Si(111) planes were used, consisting
of narrow (1/4 weight) and wide (3/4 weight) channels,
this is to be taken into account in Eq.(3) with
respective change in $d_p$ and $x_c$.
The crystal used at the Tevatron was 4 cm long with 8-mm straight ends,
having in the center $pv/R$=0.29 GeV/cm.
The theoretical dechanneling length for a straight crystal
of Si(111) is 0.646$\times$900 GeV=581 mm;
notice that for (111) it is factor of $d_p^{111}/d_p^{110}$=1.23
higher than for (110).
We take into account the dechanneling over 32 mm, as the last
8-mm part is unbent.
Eq.(3) then gives an extraction efficiency of 40.8~\%.
However, a minor correction
is discussed below.

Let us note that as the extraction efficiency is getting high,
our earlier assumption that the nuclear interactions
dominate over the crystal channeling may need correction.
To take into account the fact that the circulating particles are efficiently
removed from the ring by a crystal extraction as well,
one would require a {\em recurrent} procedure of summation:
instead of $\Sigma F_k$ one has to sum $\Sigma F^*_k$,
where $F^*_k$=$F_k(1-F^*_{k-1})$.
This ``recurrent'' correction doesn't practically affect our earlier
SPS calculations;
for Tevatron it converts 40.8\% into 34.1\%,
whereas the measured value is on the order of 30\% [13],
and the Monte Carlo simulation predicted about 35\% [3].

\section{Optimization}
For any given energy one can optimize the crystal length $L$.
In optimization we assumed the same proportion between the
bent part and the full crystal length, 3 to 4.
At 270 GeV the length used at the SPS was close to optimal,
 3.0$\pm$0.5 cm.
At 120 GeV the optimal length is 1.5$\pm$0.5 cm
resulting in the efficiency of 28\%;
at 70 GeV it is 0.8$\pm$0.2 cm with best efficiency of 38\%.
The lower energy permits a shorter crystal, and then
the multiplicity factor becomes substantial.
In Fig.2 one
can see that the analytical dependences $F_E(L)$ are very close
to those obtained earlier in Monte Carlo simulations [4].
The same maxima at the same optimal lengths are predicted.
One obvious conclusion is that the crystal extraction
experiments at the SPS and Tevatron have been working
rather far from the optimum, so there is a good possibility
for improvement.
Formula (3) predicts a good efficiency of multipass
extraction at a multi-TeV LHC, about 45\% for 0.7 mrad deflection,
with the optimal length of Si(110) crystal being 6$\pm$1 cm.
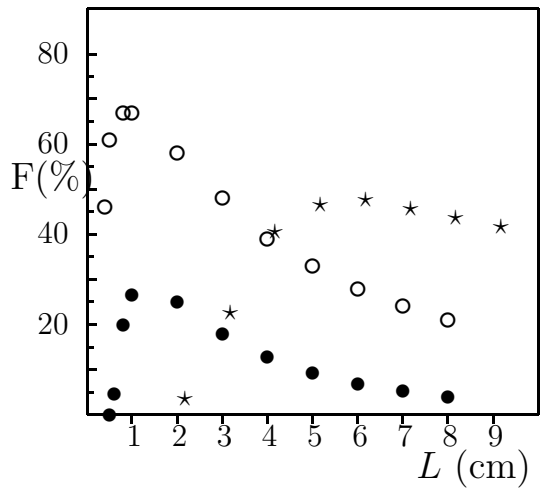
\begin{figure}[htb]
\begin{center}
\setlength{\unitlength}{.6mm}
\begin{picture}(110,100)(0,-10)
\thicklines
\linethickness{.25mm}
\put(    40.,12.8)  {\circle*{3}}
\put(    30.,17.9)  {\circle*{3}}
\put(    20.,25.)  {\circle*{3}}
\put(    10.,26.6)  {\circle*{3}}
\put(    5.,0.)  {\circle*{3}}
\put(    8.,20.)  {\circle*{3}}
\put(    6.,4.6)  {\circle*{3}}
\put(    50.,9.3)  {\circle*{3}}
\put(    60.,6.9)  {\circle*{3}}
\put(    70.,5.2)  {\circle*{3}}
\put(    80.,3.9)  {\circle*{3}}

\put(    40.,39)  {\circle {3}}
\put(    30.,48)  {\circle {3}}
\put(    20.,58)  {\circle {3}}
\put(    10.,67)  {\circle {3}}
\put(    5.,61)  {\circle {3}}
\put(    8.,67)  {\circle {3}}
\put(    4.,46)  {\circle {3}}
\put(    50.,33)  {\circle {3}}
\put(    60.,28)  {\circle {3}}
\put(    70.,24)  {\circle {3}}
\put(    80.,21)  {\circle {3}}

\put(    40.,39)  {$\star$}
\put(    30.,21)  {$\star$}
\put(    20.,2.)  {$\star$}
\put(    50.,45)  {$\star$}
\put(    60.,46)  {$\star$}
\put(    70.,44)  {$\star$}
\put(    80.,42)  {$\star$}
\put(    90.,40)  {$\star$}


\put(0,0) {\line(1,0){100}}
\put(0,0) {\line(0,1){90}}
\put(0,90) {\line(1,0){100}}
\put(100,0){\line(0,1){90}}
\multiput(10,0)(10,0){9}{\line(0,-1){2}}
\multiput(0,10)(0,10){8}{\line(1,0){2}}
\multiput(0,5)(0,5){16}{\line(1,0){1.4}}
\put(10,-7){\makebox(1,1)[b]{1}}
\put(20,-7){\makebox(1,1)[b]{2}}
\put(30,-7){\makebox(1,1)[b]{3}}
\put(40,-7){\makebox(1,1)[b]{4}}
\put(50,-7){\makebox(1,1)[b]{5}}
\put(60,-7){\makebox(1,1)[b]{6}}
\put(70,-7){\makebox(1,1)[b]{7}}
\put(80,-7){\makebox(1,1)[b]{8}}
\put(90,-7){\makebox(1,1)[b]{9}}
\put(-11,20){\makebox(1,.5)[l]{20}}
\put(-11,40){\makebox(1,.5)[l]{40}}
\put(-11,60){\makebox(1,.5)[l]{60}}
\put(-11,80){\makebox(1,.5)[l]{80}}

\put(-17,50){\large F(\%)}
\put(73,-14){\large $L$ (cm)}

\end{picture}
\end{center}
\caption{
The extraction efficiency, Eq.(3), as a function of
the crystal length $L$;  for the SPS ($\bullet$), Tevatron (o),
and Large Hadron Collider ($\star$).
}
\end{figure}

\section{New applications}
Let us mention here two interesting developments.
From Eq.(5) we see that multiplicity factor can be huge
if $L$ is very small or $L_N$ big.

\underline{\bf MeV extraction.}
One opportunity (small $L$) is inspired by the recent successful
experiment [14] on bending 3-MeV proton beam
by means of graded composition Si$_{1-x}$Ge$_x$/Si strained layers.
These epitaxial layers formed a bent crystal lattice of uniform curvature
with a thickness along the beam direction of only $L$=1 micron
(but much bigger across the beam)!
This technique allows to grow curved crystals
of any size from $\mu$m to cm's, though the bending angle
achievable is limited to few mrad's [14].
This invention allows to cover the whole spectrum of accelerator
energies (from MeV to multi-TeV) by bent crystal channeling technique.
In the context of our paper this means that one can consider
extraction from accelerators starting with MeV energies,
by crystals as short as from 1 $\mu$m.
Eqs.(1-2) predict that channeling efficiency over 99\% can be achieved
in sub-GeV (and up to several GeV) range,
thus opening a new world for bent channeling crystals
applications.
With traditional bent crystals, it was common to think that highest
efficiencies
are achievable at highest (TeV) energies as multiple scattering angles
vanish with energy faster than channeling angle does.
It's very interesting now to find that channeling efficiency is even
more boosted at lower energies due to huge multiplicity factor.
One can build a very efficient system to extract beams from
accelerators with crystals.

\underline{\bf Muon extraction.}
The other opportunity (big $L_N$) for high multiplicity factor is muons
which have formally $L_N=\infty$.
Our theory then says that factor (5) for muons is infinite,
and hence efficiency of muon channeling should be 100\%.
Actually the multiplicity factor for muons is limited by
(a) muon lifetime, (b) muon scattering out of accelerator aperture.
Quick analysis shows that the first factor dominates.
With a muon mean lifetime of 1000 turns in a 2$\times$2 TeV muon collider
[15], the typical number of encounters with crystal is $\sim$300.
This is much greater than the
corresponding quantity for protons (as limited by $L_N$).

At muon machine, the backgrounds "have the potential of killing
the concept of the muon collider"[15]; one needs a very efficient
scraping system to catch muon beam halo.
As muons cannot be absorbed, it was proposed[16] to extract
2-TeV muons with electrostatic septum as a primary element.
Surely, positive muons can be easiely steered away by bent channeling
crystals. But can we steer negative muons? Short analysis gives a very
encouraging answer.
One can channel negative particles in the same bent planes
as used for positive ones, e.g. Si(110) [17].
In same crystal Si(110), dechanneling length $L_d$ is shorter
by factor of $\sim$100 for negative particles relative to positive ones.
However, at 2 TeV $L_d$ is huge
($\sim$1 meter) for positives and modest ($\sim$1 cm) for negatives.
The required deflection angle is only 64 $\mu$rad[16]
and can be ensured by a Si crystal $\sim$1 mm long---quite shorter than $L_d$.
Let us say it simply: it is as easy to bend negatives at 64 $\mu$rad
as it is to bend positives at 6.4 mrad---which is very easy indeed!
Now let us recall that multiplicity factor greatly favors muons
again, both positive and negative.
One can build a very efficient system to handle halos at muon colliders.


\end{document}